\documentclass[runningheads]{llncs}
\usepackage{graphicx}
\usepackage{xcolor}
\usepackage{array}
\newcolumntype{L}{>{\centering\arraybackslash}m{3cm}}

\usepackage{lipsum} 

\usepackage{tikz,xcolor,hyperref}
\usepackage[open,openlevel=1]{bookmark}

\definecolor{lime}{HTML}{A6CE39}
\DeclareRobustCommand{\orcidicon}{%
	\begin{tikzpicture}
	\draw[lime, fill=lime] (0,0) 
	circle [radius=0.16] 
	node[white] {{\fontfamily{qag}\selectfont \tiny ID}};
	\draw[white, fill=white] (-0.0625,0.095) 
	circle [radius=0.007];
	\end{tikzpicture}
	\hspace{-2mm}
}

\foreach \x in {A, ..., Z}{%
	\expandafter\xdef\csname orcid\x\endcsname{\noexpand\href{https://orcid.org/\csname orcidauthor\x\endcsname}{\noexpand\orcidicon}}
}

\begin{document}
\title{Market Microstructure of Non Fungible Tokens}
%
%
\author{Mayukh Mukhopadhyay\orcidA{} \& Kaushik Ghosh\orcidB{}}
\authorrunning{Mukhopadhyay M and Ghosh K}
%
\institute{VGSOM - Indian Institute of Technology, Kharagpur\footnote[1]{{\textit{The opinions contained in this article reflect solely the personal views of the authors and do not necessarily reflect the views of the Institution, its employees or affiliates.}} \\ {Corresponding Author: \href{mailto:mayukhmukhopadhyay@iitkgp.ac.in}{\color{blue}{mayukhmukhopadhyay@iitkgp.ac.in}}}}
}
\maketitle              
\begin{abstract}
Non Fungible Token (NFT) Industry has been witnessing multi-million dollar trade in recent times. With rapid innovation of the NFT market environment by technology, innovation, and decentralization, it is becoming hard to distinguish between genuine NFT from fads and scams. This article discuss the NFT market microstructure, with a focus on price formation, market structure, transparency, and applications to other financial areas. Market manipulation in NFT market with the context of wash-sale patterns has also been surveyed. The article concludes by providing pointers on due-diligence activity that can be adopted by investors to mitigate NFT trading risk.

\keywords{Market Microstructure  \and NFT \and ERC721 \and Wash-Trading}
\textbf{JEL Classification: }{D40  \and E44  \and G14  \and G23}
\end{abstract}
\section{Introduction}
To discuss the Non fungible Tokens (NFTs) market and its microstructure, we must understand the eccentric and erratic psychology behind why we collect. No matter what we collect, physical or digital, we do so because there are a limited number of those things. Although other factors drive collectors to collect, such as investment, speculation, popularity, emotional connection and the fear of missing out (FOMO), at the core of collecting is scarcity. NFTs are unique and rare digital collectibles which are secured over a blockchain. An NFT provides authenticity of origin, ownership, uniqueness (scarcity), and permanence for any particular item. On the other hand, Market microstructure analyze the process by which investors’ latent demands are translated into executed trades~\cite{ref_article5}.

The article has been organized into four major sections. The first section deals with various aspects of price formation and discovery of NFTs. The second section discusses the design and structure of NFT marketplace. The third section emphasizes on the information available to market participants and how manipulation occurs in the NFT market. In the final section we discuss various aspects of NFT market in legal and corporate due-diligence activities where we use the Howey test to determine whether NFTs should be treated as securities or asset-class.

\section{Price Formation and Discovery}

Price formation is the process by which prices confine new information. It is a process of gathering information to ensure market participants know enough about the prices of the NFT being traded in the marketplace, so that they can make well-informed decisions. NFTs are generally traded in two ways - Fixed Price and Timed Auction. In the fixed price formation, the seller fixes the price of the NFT for a certain period of time, generally ranging from 7 to 180 days. The interested buyer can directly purchase the NFT by transfering the set-price in USD or ETH equivalent amount. In case the buyer feels that the price of NFT is overestimated, then the buyer can place an offer price lower than the fixed amount. However, the seller is not obligated to sell at the lower price and can wait till the time period expires or a new offer is received from another prospective buyer with a higher offer price. Timed Auction are of many types, but in NFT marketplace like Opensea~\cite{ref_article3}, two kind of timed auction can be opted by the seller - English Auction and Dutch Auction. In English auction, the highest bidder wins at the end of the timed auction. Any bid made in the last 10 minutes of the auction extends each auction by another 10 minutes. The seller can set a minimum bid amount at the start of the timed auction. In the Dutch Auction, NFT price keeps falling till someone purchases it. For example, an NFT listing can be time-auctioned for 5 days where the price starts at 5 ETH and ends at 1 ETH after a duration of 5 days.

Unlike Bitcoin or Ethereum, the maximum supply of any NFT is 1. So, NFT are traded as assets rather than securities. Although, whether cryptocurrency should be traded as asset or as security is in itself a highly debated topic in the financial regulation community, in this article we can assume NFT to be a digital asset due to its scarcity of supply being just 1 i.e. non-fungible. Hence Garman's Logic of market maker's inventory~\cite{ref_article5} would not work for NFT. However, Inventory is not the only consideration for a dealer. The market maker loses to informed traders, on average, but recoups the losses on trades with liquidity motivated traders. Bid–ask spreads contain a component attributable to asymmetrical information. French and Roll (1986) found that, on an hourly basis, the variance during trading periods is 20 times larger than the variance during nontrading periods~\cite{ref_article5}.This suggests that trading itself is an important source of volatility; for markets to be efficient, someone has to make them efficient. The dark side of this is market-manipulation by wash-trades\footnote[1]{\textbf{wash-trade} \textit{A wash trade or wash sale occurs when an investor sells or trades a token at a loss, and within 30 days before or after, buys another one that is substantially similar. It also happens if the individual sells the token at a loss, and their dependent or an institution they control buys a substantially similar token within 30 days.}} of NFTs that we survey in later section.

Price discovery is the culmination of market sentiment from buyer and seller at a fixed point in time. This is the price at which an NFT changes its ownership i.e. gets traded at market order price. While price formation of NFT is analogous to double auction market, the price discovery is not like the single auction market.Price discovery at NFT marketplace is currently hype-driven with a combination of legacy bias of investors. 

Two vital components of NFT marketplace, that are not found in vanilla securities trading, are secondary market royalties to original creator and gas fee. Gas fee is something different from institutional commission as it is a fee that is not consumed by the marketplace for facilitating NFT trade but serves as an inherent mining fee for actually making the decentralized transaction between two parties. Platforms like polygon network has been collaborating with marketplaces to nullify this gas fee to miners but has a limited reach till date.

\section{Market Structure and Design}
Market structure affects the speed and quality of price discovery, liquidity, and the cost of trading. What ultimately matters is not the medium of communication between the investor and the market but the protocols that translate the order into a realized transaction~\cite{ref_article4}. From the design perspective, sustainability of the NFT market and identification of the health indicators are more important than the volume of trade occuring in the NFT marketplace. 

With the evolution of NFT market, we can identify three kinds of players - active wallets, buyer, and seller. Active wallet means those entities or person who have interacted with NFT smart contract. Although, an individual can possess more than one active wallet, the trend remains insignificant.  Buyers are the number of wallets that bought atleast one NFT across the year. Sellers are the number of wallets that sold atleast one NFT across the year. Traded in US Dollar  (TUSD) represents the total volume traded in the NFT market which is an important parameter of market sustainability. 

TUSD can be divided into two classes - USD from sales (USDS) and USD from dApps (USDD). USDS represents the value that was transferred through a transaction between a buyer and a seller. This volume includes primary sales, from a project or artist to the buyer, and secondary market sales, between players or collectors. USDD represents the trade volume that includes all interactions with Smart Contracts including financial. This can correspond to the improvement of an asset, its modification, breeding and the creation of asset groups. The USDD is more representative of the activity within the projects. As the USDD increases and widens the gap from USDS, the NFT industry, as a whole, move more towards mature industry and player engagement and away from hype-driven speculative store-of-value.

\subsection{NFT Marketplace}

\begin{figure}
\includegraphics[width=\textwidth]{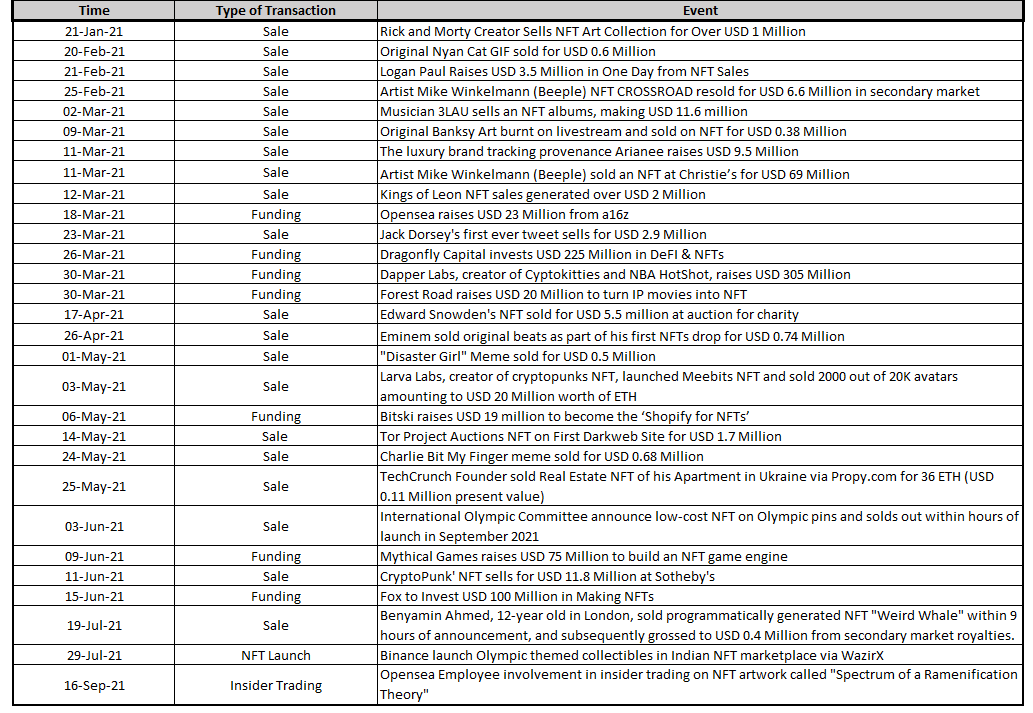}
\caption{Major NFT Trading Events from January to September 2021 } \label{tabfig1}
\end{figure}

NFT marketplace can be categorized into two levels. One is at project level and another is at ownership level. At project level, NFT market can be segregated into 6 main segments - Art, Collectibles, Sports, Utility, Metaverse, and Games. At ownership level, the NFT market is segmented into Primary and Secondary Market. Leading platforms like Opensea.io, Rarible.com, WazirX, Binance, and Mintable.app facilitates players to mint, trade, buy, and auction NFTs via metamask wallet account that links the platform and cryptocurrency exchanges with fiat currency banking entities. Art and collectibles captures the major market segments with large volume of trades and major events in 2021 (ref. Figure~\ref{tabfig1}). Opeasea is a major platform facilitating such transaction both in primary and secondary ownership market~\cite{ref_article7}. A nonexistent secondary market may indicate that owners HODL\footnote[2]{\textit{Hold on for dear life or \textbf{HODL} is an investor behaviour where people do not impulsively sell a token when its price drops or rises dramatically.}} their assets, waiting for the opportune moment to sell. Of course, on the other hand it can also reflect an abandoned community. The primary market will only provide information on the issuance of new Tokens by the project.

\section{Information} 

Market microstructure depends on the dynamics between two kinds of information - Transparency and Disclosure. Transparency deals with the availability of market information to the market participants~\cite{ref_article9}, while disclosure deals with the impact on participants behaviour due to the availabity of market information. Transparency has two dimensions - Pretrade and Posttrade. Pretrade transparency refers to the wide dissemination of current bid-ask quotations, and existence of large order imbalances~\cite{ref_article1}. Posttrade transparency refers to the public and timely disclosure of information on past trades. In a totally automated trading system, where the components of order flow cannot be distinguished, transparency is not an issue~\cite{ref_article5}. There is broad agreement among market researchers that both pre- and post-trade transparency matter and both affect liquidity and price efficiency~\cite{ref_article4,ref_article9}. Second, greater transparency, both pre- and posttrade, is generally associated with more informative prices~\cite{ref_article1,ref_article9}. Third, complete transparency is not always \textit{beneficial} to the operation of the market~\cite{ref_article1,ref_article5,ref_article9}. 

\subsection{Market Manipulation}

\begin{figure}
\includegraphics[width=\textwidth]{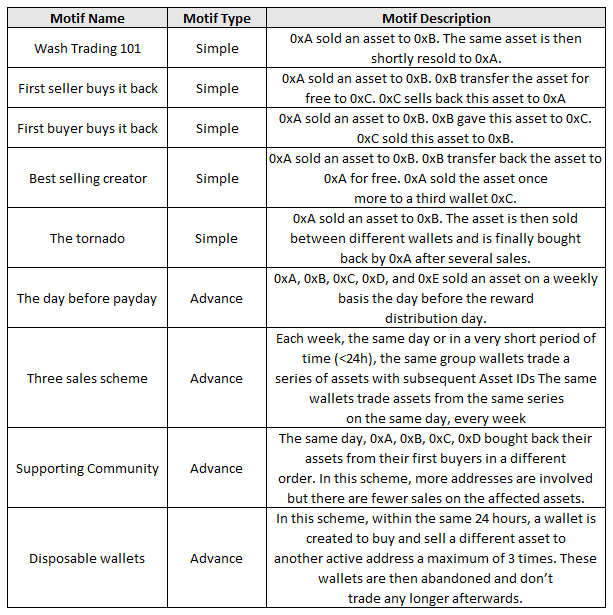}
\caption{Summary of Wash-Trade Motifs~\cite{ref_article8} } \label{tabfig2}
\end{figure}

To quantify market manipulation using Wash Trading, it is first necessary to identify in what form Wash Trading takes place, how to recognize it and what reasonable doubts leads to consider a transaction is an illegitimate one. NonFungible Corporation, established in 2018, has been tracking such illicit trade behaviour on public Ethereum Blockchain and have identified many simple and advanced patterns~\cite{ref_article8}. The motifs for Wash Trading have evolved with the development of the NFT ecosystem. Figure~\ref{tabfig2} summarizes all the motifs of wash-trade pattern. The main signature of identifying patterns are of two kinds - \textit{abnormal} transaction activity and \textit{recurring} transaction activity. 

Wash Trading is necessarily done on purpose. Manipulating a market involves generating a large volume of transactions, which incurs costs and can be very expensive. A Market in full development and underpinned by such economic issues is likely to attract speculation and market bad behavior. The actors practicing such behaviour may have different motivations but the greater the lure of profit, the more the techniques will get evolved. Four key reasons that motivates Wash-Trade~\cite{ref_article8}:
\begin{enumerate}
  \item \textbf{Market Making - } Artificially inflates the price of an asset or asset type.
  \item \textbf{Rate Making - } Artificially
increases the value of Artist to sell future assets at a higher price.
  \item \textbf{Incentivizing - } On platforms that reward for each trade, reward is amount is substantially more than wash-trading cost.
  \item \textbf{Project Promotion - } Trading volume improves ranking of a project to attract organic investors.
\end{enumerate}

The question of regulation therefore arises to prevent these practices but the decentralized nature of a public blockchain makes it impossible to actually regulate these behaviors. Whistleblowing, raising awareness, and due-diligence are the best way to empower NFT ecosystem~\cite{ref_article8}. Banning of participants will only shift the focus of the problem. Consensus and compromise provide a middle ground to avoid tight regulation and eventual centralization in the NFT marketplace.

\section{Applications} 

An interesting application of market microstructure in the NFT-pricing area concerns
technical analysis, in which past price movements are used to predict future returns~\cite{ref_article4}. Technical analysis might help traders discover hidden liquidity, over short horizons~\cite{ref_article5}. Due-diligence and understanding various legal aspects can further help in understanding the NFT-pricing dynamics.

\subsection{Due-Diligence} 

As the NFT ecosystem is already extremely broad, from Art to Gaming and Metaverses, the due diligence process is necessarily different depending on the particular asset type. Five basic questions can serve as an initial checklist for identifying authentic NFT projects from scams and fads.

\begin{enumerate}
  \item Is the project verified?
  \item Who is the Team behind the project?
  \item What makes the asset valuable?
  \item Is the project truly decentralized?
  \item Is there a true scarcity?
\end{enumerate}

\subsection{Laws, Rights, and Taxes} 

With any new use of technology, it usually takes a while for regulatory agencies, legislatures, and the courts to catch up to the technology’s rapid adoption. No specific legal doctrine has yet been set for NFTs, but we can deduce how the law may be applied by looking more generally at how the law is applied to cryptocurrencies, art, and collectibles.

In 1946, the Supreme Court of the United States heard the case of \textit{Securities and Exchange Commission (SEC) v. W. J. Howey Co.}, which involved whether a specific leaseback agreement should be considered an investment agreement~\cite{ref_article10}. If so, that would make it a security and subject to SEC regulations. In its landmark decision, the Supreme Court outlined four factors for determining whether an investment is a security, that became the Howey test: 

\begin{enumerate}
  \item An investment of money
  \item In a common enterprise
  \item With the expectation of profit
  \item To be derived from the efforts of a promoter or other third party
\end{enumerate}

Although, the SEC has not issued any guidance on NFTs yet, we can utilize the Howey test to determine the legal nature of NFTs:
\begin{enumerate}
  \item Purchasers of NFTs invest money or cryptocurrency (something of value).
  \item There generally do not seem to be common enterprises associated with NFTs. Instead, most NFTs are one-offs or limited editions of digital art, are collectibles, or have some utility, such as an in-game item.
  \item Some people may purchase NFTs as an investment, with an expectation of profit, while others purchase NFTs for their subject matter and for building a collection.
  \item Generally, there’s no third party promoting the value of NFTs that have been sold.
\end{enumerate}

NFTs being non-fungible, are more akin to works of art or collectibles, which are not securities, than to fungible cryptocurrencies. If an NFT has a massive supply or a vast number of editions, it leans more toward a fungible token, and the line becomes less clear. It could even lean toward securities for certain fractional NFTs. Fractional NFTs are tokens that represent a fractional ownership of an NFT. Mutual funds work the same way too. Investors pool money into a fund, and that money is used to buy assets. Each investor owns a fraction of the entirety of the fund. Also, when it comes to fractionalized art ownership, while companies have different business models, most file with the SEC.

Intellectual property (IP) is a property that is derived from creativity. IP encompasses copyright, trademark, patent, and trade secrets. In NFT space, copyright and trademark has the central focus over other IP classes.

A copyright is created once the work is fixed in a tangible medium. In other words, the work can’t just be in our head, spoken, sung, or performed, unless the latter three were recorded. The copyright is separate and distinct from the actual work. The copyright is the intangible right pertaining to the tangible work that is endowed upon the creator of the work. When we purchase an NFT, we are not purchasing the copyright in that NFT. The creator retains the copyright.We do have the right to sell the NFT at any time. Copyrights can potentially be the main content of an NFT. This type of \textit{Copyrights-NFT}, a participation in an income stream, seems like an investment, and might be considered a security by the SEC.

A \textit{trademark} is generally a symbol, design, word, or phrase (or combination thereof) that identifies the source of goods (products). A \textit{service-mark} is similar to a trademark, but it identifies a service, not a product. Generally,\textit{trademark} is a broad term that encompasses both trademarks and service marks.The main crux of trademark law is preventing consumer confusion regarding the origin of goods. When it comes to creating NFTs, it’s generally OK to use a company’s trademark for commentary, criticism, and parody purposes. Unfortunately, there’s no bright line test to determine whether our use of a trademark is appropriate.

NFTs are not immune from taxes. If we’re creating and selling NFTs, we’ll be responsible for paying taxes on our income from those sales. However, we should be able to deduct the expenses that we incurred in connection with our creation, minting, listing, and promotion of our NFTs. NFTs are also subject to capital gains tax. If we sell an NFT that we held for less than one year, that would be considered short-term capital gains, and the tax we owe would be based on our regular income tax rate for both federal and state taxes. If we sell an NFT that we held for at least one year, that would be considered long-term capital gains, and the federal tax we owe would be based on the capital gains tax rate, which is generally more advantageous than our regular income rate. We are subjected to capital gains tax when we purchase the NFT with cryptocurrency. Sales tax on NFTs represents a tricky situation. Generally, sales taxes are applied to the sale of goods and services by the state or federal government in which the sale occurred. With the sale of NFTs, the first question would be, where did the sale occur? Apparently, some states' law define digital items as \textit{items that were downloaded}. Since NFTs remain on blockchain and are not downloaded, are NFTs a digital item in legal context? In such cases, sales tax cannot be levied on NFT sales~\cite{ref_article2}. What if the NFT contains perks that are physical items or services? For example, cryptokicks are patented NFTs by NIKE that represent a physical shoe in the real world~\cite{ref_article6}. Sales tax, in this case, might be applied to the value of the perks.

\section{Conclusion and Future of NFT} 

To conclude, NFT markets are a great deal more complex than commonly believed.\textit{One size fits all} approaches to regulation
and policy making should be avoided. NFTs have a bright future because they’re providing a bridge to digital economies~\cite{ref_article2}.

\begin{figure}
\includegraphics[width=\textwidth]{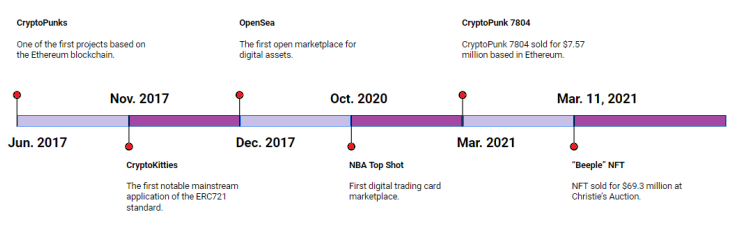}
\caption{Timeline of NFTs~\cite{ref_article7} } \label{tabfig3}
\end{figure}

Nobody knows what will become the most prominent use of NFTs. Starting from an idea in 2017 as ERC721, NFT has become an exotic medium of digital art where Auction houses like Christie are participating in muti-million dollar NFT sales (ref. Figure~\ref{tabfig3}). Looking at NFTs solely as a speculative art asset is narrow-minded and misses the multitude of future uses of
NFTs.

The NFT-ification of everything will take place over the next decade. Monetization of corporate legacy, tokenization of end of life-cycle (EOSL) digital products, and catalyzing digital circular economy, the applications of NFT remains endless and uncharted. It
is up to us to build an industry capable of changing the world and revolutionizing uses and
economic models.

\end{document}